\documentclass[prd,showpacs, showkeys]{revtex4}
\begin{document}
\newcommand{\dfrac}[2]{\frac{\displaystyle #1}{\displaystyle #2}}
\baselineskip=12pt
\title{ Cornwall-Jackiw-Tomboulis
effective potential for quark propagator in real-time thermal
field theory and Landau gauge\footnote{The project supported by
the National Natural Science Foundation of China.} \\
}
\author{Zhou Bang-Rong}
\affiliation{College of Physical Sciences, Graduate School of the
Chinese Academy of Sciences, Beijing 100049, China} \affiliation{
CCAST (World Laboratory), P.O.Box 8730, Beijing 100080, China }
\date{March 30, 2005}
\begin{abstract}
We complete the derivation of the Cornwall-Jackiw-Tomboulis
effective potential for quark propagator at finite temperature and
finite quark chemical potential in the real-time formalism of
thermal field theory and in Landau gauge. In the approximation
that the function $A(p^2)$ in inverse quark propagator is replaced
by unity, by means of the running gauge coupling and the quark
mass function invariant under the renormalization group in zero
temperature Quantum Chromadynamics (QCD), we obtain a calculable
expression for the thermal effective potential which will be a
useful means to research chiral phase transition in QCD in the
real-time formalism.
\end{abstract}
\pacs{11.10.Wx; 12.38.Lg; 11.30.Rd; 12.38.Aw} \keywords{CJT
effective potential, real-time thermal field theory, QCD,
 quark propagator, Landau gauge}
 \maketitle
Effective potential \cite{kn:1,kn:2,kn:3,kn:4,kn:5} is a basic
means to research the vacuum of quantum field theory and
spontaneous symmetry breaking. For discussion of chiral symmetry
phase transition in QCD, one must generalize the
Cornwall-Jackiw-Tomboulis (CJT) effective potential \cite{kn:6}
for quark propagator to finite temperature and finite quark
chemical potential case. At present, some work on the thermal  CJT
effective potential are mainly based on the imaginary-time
formalism \cite{kn:7}, very few were found in literature which are
based on the real-time formalism \cite{kn:8}. In fact, in the
real-time formalism, besides that it is no longer necessary to do
analytic continuation of the discrete Matsubara frequency, one can
directly use the well-known results in zero temperature QCD, e.g.
the running gauge coupling and the running quark mass function
consistent with the  renormalization group (RG) analysis before
thermalizing propagators by thermal transformation matrices, and
this will certainly bring great advantages for practical
calculation.\\
\indent In a preceding paper \cite{kn:9}, for deriving the thermal
Schwinger-Dyson (SD) equation for quark self-energy, we gave a
formal expression for the thermal CJT effective potential for
quark propagator in Landau gauge. In this paper, we will continue
to reduce the formal expression to an explicit formula for future
use. \\
\indent For convenience of the derivation and self-containment of
the paper, we will first put down the relevant results given in
Ref. \cite{kn:9}.  The formal expression for the CJT effective
potential for quark propagator at finite temperature $T$ and
finite quark chemical potential $\mu$ has been expressed by
\begin{equation}
V[G,G^*]=V_1[G,G^*]+V_2[G,G^*]
\end{equation}
where in momentum space the contribution from one-loop diagrams
\begin{eqnarray}
V_1[G,G^*]&=& iN_fN_c\Big(
\big\langle\cos^2\theta_p\big\{\textrm{tr}\ln\big[S(p)G^{-1}(p)\big]+
\textrm{tr}\big[S^{-1}(p)G(p)\big]\big\}
\big\rangle\nonumber \\
  &&+\big\langle\sin^2\theta_p\big\{\textrm{tr}\ln\big[S^*(p)G^{*-1}(p)\big]+
  \textrm{tr}\big[S^{*-1}(p)G^*(p)\big]\big\}\big\rangle
  -\big\langle\textrm{tr}1\big\rangle\Big),
\end{eqnarray}
and the contribution from two-loop vacuum diagrams
\begin{eqnarray}
 V_2[G,G^*]&=& \frac{i}{2}N_fN_cC_2(\underline{N}_c) \big\langle\big\langle
 g^2\big[(p-q)^2\big]\textrm{tr}
 \big[\gamma^{\mu}G^{11}_T(p)\gamma^{\nu}G^{11}_T(q)\big]
 \big[D_{\mu\nu}(p-q)\big]^{11}_T\big\rangle\big\rangle \nonumber\\
  && -\frac{i}{2}N_fN_cC_2(\underline{N}_c) \big\langle\big\langle g^2\big[(p-q)^2\big]\textrm{tr}
 \big[\gamma^{\mu}G^{12}_T(p)\gamma^{\nu}G^{21}_T(q)\big]
 \big[D_{\mu\nu}(p-q)\big]^{12}_T\big\rangle\big\rangle.
\end{eqnarray}
In the above denotations, $\langle\cdots\rangle$ and
$\langle\langle\cdots\rangle\rangle$ represent respectively the
integration $\int d^4p/(2\pi)^4$ and $\int d^4pd^4q/(2\pi)^8$,
$N_f$ and $N_c$ are respectively the number of flavor and color of
the quarks, $C_2(\underline{N}_c)=(N_c^2-1)/2N_c$,
\begin{equation}
S(p)=i/(\not\!{p}+i\varepsilon)\;\; \mbox{and}\;
\;S^*(p)=-i/(\not\!{p}-i\varepsilon)\;\;\mbox{with}\;\;\not\!{p}=\gamma^{\mu}p_{\mu}
\end{equation}
are the propagator of massless quark and its conjugate,
\begin{equation}
G(p)=i/\big[\not\!{p}-m(p^2)+i\varepsilon\big]\;\;\mbox{and}\;\;
G^*(p)=-i/\big[\not\!{p}-m(p^2)-i\varepsilon\big]
\end{equation}
are complete quark propagator and its conjugate, in which we have
made the approximation $A(p^2)=1$ on the basis of the arguments
given in Sect.IV of Ref. \cite{kn:9} and the replacement
$B(p^2)\rightarrow m(p^2)$, $g^2\big[(p-q)^2\big]$ is the running
gauge coupling, $\textrm{tr}$ means the trace of spinor matrices,
\begin{equation}
\sin^2\theta_p=\theta(p^0)\tilde{n}(p^0-\mu)+\theta(-p^0)\tilde{n}(-p^0+\mu)\;\;\mbox{with}\;\;
\tilde{n}(p^0-\mu)=1/\big[e^{\beta(p^0-\mu)}+1\big]\;\;\mbox{and}\;\;\beta=1/T,
\end{equation}
\begin{equation}
G^{11}_T(p)=\cos^2\theta_pG(p)-\sin^2\theta_pG^*(p),\;\;
G^{12}_T(p)=-\cos\theta_p\;\sin\theta_p\;
e^{\beta\mu/2}\big[G(p)+G^*(p)\big]=-e^{\beta\mu}G_T^{21}(p),
\end{equation}
and
\begin{equation}
\big[D_{\mu\nu}(p-q)\big]^{11}_T=D_{\mu\nu}(p-q)-g_{\mu\nu}2\pi
n(p^0-q^0)\delta\big[(p-q)^2\big],\;\;\big[D_{\mu\nu}(p-q)\big]^{12}_T=
-g_{\mu\nu}e^{\beta|p^0-q^0|/2}2\pi
n(p^0-q^0)\delta\big[(p-q)^2\big]
\end{equation}
with
\begin{equation}
D_{\mu\nu}(p-q)=\frac{-i}{(p-q)^2+i\varepsilon}\Delta_{\mu\nu}(p-q),\;\;
\Delta_{\mu\nu}(p-q)=
g_{\mu\nu}-\frac{(p-q)_{\mu}(p-q)_{\nu}}{(p-q)^2+i\varepsilon}
\end{equation}
and
\begin{equation}
n(p^0-q^0)=1/\big[e^{\beta|p^0-q^0|}-1\big].
\end{equation}
\indent Now we begin performing detailed calculations of
$V[G,G^*]$. First rewrite $V_1[G,G^*]$ in Eq.(2) by
\begin{equation}
V_1[G,G^*]=V_{10}+V_{1T}
\end{equation}
with
\begin{equation}
V_{10}= iN_fN_c\left(
\Big\langle\textrm{tr}\ln\big[S(p)G^{-1}(p)\big]+\textrm{tr}\big[S^{-1}(p)G(p)\big]
-\textrm{tr}1\Big\rangle\right)
\end{equation}
and
\begin{equation}
V_{1T}=iN_fN_c\Big\langle\sin^2\theta_p\big\{\textrm{tr}\ln\big[S^*(p)G^{*-1}(p)\big]+
  \textrm{tr}\big[S^{*-1}(p)G^*(p)\big]-\textrm{tr}\ln\big[S(p)G^{-1}(p)\big]-
  \textrm{tr}\big[S^{-1}(p)G(p)\big]\big\}\Big\rangle.
\end{equation}
By means of Eqs.(4) and (5) we get
\begin{eqnarray}
 V_{10}&=&iN_fN_c\Big\langle\textrm{tr}
 \left\{\ln\left[1-\frac{m(p^2)}{\not\!{p}+i\varepsilon}\right]+
 \frac{m(p^2)}{\not\!{p}-m(p^2)+i\varepsilon}\right\}\Big\rangle\nonumber \\
 &=& iN_fN_c\Big\langle\textrm{tr}
     \left\{\sum_{k=1}^{\infty}\frac{-1}{k}\left[\frac{m(p^2)}{\not\!{p}+i\varepsilon}\right]^k+
     \frac{m(p^2)}{\not\!{p}-m(p^2)+i\varepsilon}\right\}\Big\rangle\nonumber\\
  &=& i2N_fN_c\int\frac{d^4p}{(2\pi)^4}
 \left\{\ln\left[1-\frac{m^2(p^2)}{p^2+i\varepsilon}\right]+
 \frac{2m^2(p^2)}{p^2-m^2(p^2)+i\varepsilon}\right\}\nonumber \\
   &=&-2N_fN_c\int\frac{d^4\bar{p}}{(2\pi)^4}
 \left\{\ln\frac{\bar{p}^2+m^2(\bar{p}^2)}{\bar{p}^2}-
 \frac{2m^2(\bar{p}^2)}{\bar{p}^2+m^2(\bar{p}^2)}\right\}
\end{eqnarray}
In the last line we have made the Wick rotation and the changes of
variables
\begin{equation}
\bar{p}^0=ip^0, \; \bar{p}^i=p^i \;(i=1,2,3).
\end{equation}
In similar way we obtain
\begin{eqnarray}
  V_{1T}&=& i2N_fN_c\Big\langle\sin^2\theta_p
  \left\{\ln\left[1+\frac{m^2(p^2)}{p^2-m^2(p^2)+i\varepsilon}\right]-
  \ln\left[1+\frac{m^2(p^2)}{p^2-m^2(p^2)-i\varepsilon}\right]\right.\nonumber \\
  &&\left. -2p^2\left[\frac{1}{p^2-m^2(p^2)+i\varepsilon}-\frac{1}{p^2-m^2(p^2)-i\varepsilon}
  \right]\right\}\Big\rangle\nonumber \\
  &=& -4N_fN_c\Big\langle\sin^2\theta_p\textrm{Im}
  \left\{\ln\left[1+\frac{m^2(p^2)}{p^2-m^2(p^2)+i\varepsilon}\right]-
  \frac{2p^2}{p^2-m^2(p^2)+i\varepsilon}\right\}\Big\rangle\nonumber \\
  &=& -4N_fN_c\Big\langle\sin^2\theta_p\bigg\{
  -\pi m^2(p^2)\delta\big[p^2-m^2(p^2)\big]\sum_{k=1}^{\infty}
  \left(\frac{-m^2(p^2)\big[p^2-m^2(p^2)\big]}{\big[p^2-m^2(p^2)\big]^2+\varepsilon^2}\right)^{k-1}+
  2\pi p^2\delta\big[p^2-m^2(p^2)\big]\bigg\}\Big\rangle\nonumber\\
  &=&-4\pi N_fN_c\Big\langle\sin^2\theta_p
  m^2(p^2)\delta\big[p^2-m^2(p^2)\big]\Big\rangle,
\end{eqnarray}
where the relation
$\delta\big[p^2-m^2(p^2)\big]\big[p^2-m^2(p^2)\big]=0$ has been used.\\
\indent Next we rewrite $V_2[G,G^*]$ in Eq.(3) in terms of
Eqs.(7)-(9) by
\begin{equation}
V_2[G,G^*]=V_{20}+V_{2T},
\end{equation}
where $V_{20}$ and $V_{2T}$ are respectively the contributions
from the two-loop vacuum diagrams without and with thermal
corrections of propagators. $V_{20}$ can be expressed by
\begin{eqnarray}
  V_{20}&=& -\frac{1}{2}N_fN_cC_2(\underline{N}_c)\Big\langle\Big\langle g^2\big[(p-q)^2\big]
  \textrm{tr}\left\{\gamma^{\mu}\big[\not\!{p}+m(p^2)\big]\gamma^{\nu}
  \big[\not\!{q}+m(q^2)\big]\right\}
  \frac{1}{p^2-m^2(p^2)+i\varepsilon}\cdot\frac{1}{q^2-m^2(q^2)+i\varepsilon}\nonumber  \\
  && \cdot\frac{1}{(p-q)^2+i\varepsilon}\Delta_{\mu\nu}(p-q)\Big\rangle\Big\rangle\nonumber \\
  &=&-2N_fN_cC_2(\underline{N}_c)\Big\langle\Big\langle g^2\big[(p-q)^2\big]
  \big[3m(p^2)m(q^2)+(p-q)^2 E(p,q)\big]\frac{1}{p^2-m^2(p^2)+i\varepsilon}
  \cdot\frac{1}{q^2-m^2(q^2)+i\varepsilon}\nonumber \\
  &&\cdot\frac{1}{(p-q)^2+i\varepsilon}\Big\rangle\Big\rangle\nonumber\\
  &=&-6N_fN_cC_2(\underline{N}_c)\int\frac{d^4\bar{p} d^4\bar{q}}{(2\pi)^8}
  g^2\big[\textrm{max}(\bar{p}^2,\bar{q}^2)\big]
  \frac{m(\bar{p}^2)}{\bar{p}^2+m^2(\bar{p}^2)}\cdot\frac{m(\bar{q}^2)}{\bar{q}^2+m^2(\bar{q}^2)}
  \cdot\frac{1}{(\bar{p}-\bar{q})^2}
\end{eqnarray}
where we have used the result
\begin{equation}
\textrm{tr}\Big\{\gamma^{\mu}\big[\not\!{p}+m(p^2)\big]\gamma^{\nu}\big[\not\!{q}+m(q^2)\big]
\Big\}\Delta_{\mu\nu}(p-q)=4\big[3m(p^2)m(q^2)+(p-q)^2 E(p,q)\big]
\end{equation}
with the denotation
\begin{equation}
E(p,q)=1-\frac{1}{2}\left[p^2+q^2+\frac{(p^2-q^2)^2}{(p-q)^2}\right]/(p-q)^2,
\end{equation}
and after the Wick rotation, the running gauge coupling
$g^2\big[(\bar{p}-\bar{q})^2\big]$ has been approximated by
\cite{kn:10}
\begin{equation}
g^2\big[\textrm{max}(\bar{p}^2,\bar{q}^2)\big]=\theta(\bar{p}-\bar{q})g^2(\bar{p}^2)+
\theta(\bar{q}-\bar{p})g^2(\bar{q}^2)
\end{equation}
so that
\begin{equation}
\int d\Omega_{\bar{q}} E(\bar{p},\bar{q})=0.
\end{equation}
By using Eq.(19) and the result that
\begin{equation}
\textrm{tr}\Big\{\gamma^{\mu}\big[\not\!{p}+m(p^2)\big]\gamma^{\nu}\big[\not\!{q}+m(q^2)\big]\Big\}
g_{\mu\nu}=4\;\big[4m(p^2)m(q^2)-2p\cdot q\big]
\end{equation}
and considering some symmetry of the integrand under the exchange
between the variables $p$ and $q$ we may express $V_{2T}$ by
\begin{eqnarray}
 V_{2T}&=&i4\pi N_fN_cC_2(\underline{N}_c)g^2(0)\Big\langle\Big\langle
 \big[4m(p^2)m(q^2)-p^2-q^2\big]
 n(p^0-q^0)\delta\big[(p-q)^2\big]\nonumber \\
 && \left\{\frac{1}{p^2-m^2(p^2)+i\varepsilon}\cdot\frac{1}{q^2-m^2(q^2)+i\varepsilon}+
 i4\pi\sin^2\theta_p\delta\big[p^2-m^2(p^2)\big]\textrm{P}\frac{1}{q^2-m^2(q^2)}\right\}
 \Big\rangle\Big\rangle\nonumber \\
 &&-8\pi N_fN_cC_2(\underline{N}_c)\Big\langle\Big\langle\frac{g^2\big[(p-q)^2\big]}{(p-q)^2+i\varepsilon}
 \big[3m(p^2)m(q^2)+(p-q)^2 E(p,q)\big]\sin^2\theta_p\delta\big[p^2-m^2(p^2)\big]\nonumber \\
  &&\left\{\frac{i}{q^2-m^2(q^2)+i\varepsilon}-\pi
  \sin^2\theta_q\delta\big[q^2-m^2(q^2)\big]\right\}\Big\rangle\Big\rangle
  +\Delta V_{2T},
\end{eqnarray}
where "P" in the second line of Eq.(24) means the principle value
of  the integration and
\begin{eqnarray}
\Delta
V_{2T}&=&-i16\pi^3N_fN_cC_2(\underline{N}_c)\Big\langle\Big\langle
g^2\big[(p-q)^2\big]\big[4m(p^2)m(q^2)-p^2-q^2\big]
 \delta\big[p^2-m^2(p^2)\big]\delta\big[q^2-m^2(q^2)\big]\delta\big[(p-q)^2\big]\nonumber\\
 && n(p^0-q^0)\left\{\sin^2\theta_p\sin^2\theta_q-\sin^2\theta_p+
 \sin\theta_p\cos\theta_p\sin\theta_q\cos\theta_qe^{\beta|p^0-q^0|/2}\right\}
 \Big\rangle\Big\rangle
\end{eqnarray}
Let $p^2=m_1^2$ is a real root of the equation $p^2=m^2(p^2)$,
i.e. $m_1$ obeys the equation $m_1^2=m^2(m_1^2)$, then we will
have $\delta\big[p^2-m^2(p^2)\big]=\delta(p^2-m_1^2)/f(m_1^2)$
with $f(m_1^2)=|1-\partial m^2(p^2)/\partial p^2|_{p^2=m_1^2}$.
Thus we get
\begin{eqnarray}
 \Delta V_{2T}&=&-i32\pi^3N_fN_cC_2(\underline{N}_c)g^2(0)\frac{m_1^2}{f^2(m_1^2)}
 \Big\langle\Big\langle
 \delta(p^2-m_1^2)\delta(q^2-m_1^2)\delta\big[(p-q)^2\big]\nonumber\\
 && n(p^0-q^0)\left\{\sin^2\theta_p\sin^2\theta_q-\sin^2\theta_p+
 \sin\theta_p\cos\theta_p\sin\theta_q\cos\theta_qe^{\beta|p^0-q^0|/2}\right\}
 \Big\rangle\Big\rangle
 \nonumber \\
  &=& -i32\pi^3N_fN_cC_2(\underline{N}_c)g^2(0)\frac{m_1^2}{f^2(m_1^2)}
  \Big\langle\Big\langle
 \delta\big[(p+q)^2-m_1^2\big]\delta(q^2-m_1^2)\delta(p^2)\nonumber\\
 && n(p^0)\left\{\sin^2\theta_{p+q}\sin^2\theta_q-\sin^2\theta_{p+q}+
 \sin\theta_{p+q}\cos\theta_{p+q}\sin\theta_q\cos\theta_qe^{\beta|p^0|/2}\right\}
 \Big\rangle\Big\rangle
\end{eqnarray}
In the second equality above, we have made the change of variables
$p-q\rightarrow p$.  It is indicated that the condition in which
the three $\delta$ functions in Eq.(26) are not equal to zeroes
simultaneously is $p^0=|\vec{p}|=0$. Taking this and the
definition of $\sin^2\theta_p$ given by Eq.(6) into account, we
are led to that
$$\big[\sin^2\theta_{p+q}\sin^2\theta_q-\sin^2\theta_{p+q}+
 \sin\theta_{p+q}\cos\theta_{p+q}\sin\theta_q\cos\theta_qe^{\beta|p^0|/2}\big]\big|_{p^0=0}=
\sin^2\theta_q\big[\sin^2\theta_q-1+\cos^2\theta_q\big]=0.$$ On
the other hand, it is also noted that because
$$\int d^4p\delta(p^2)n(p^0)=\int dp^0\int d\Omega_{p\vec{}}\int
d|\vec{p}||\vec{p}|^2\frac{1}{2|\vec{p}|}\big[\delta(p^0-|\vec{p}|)+\delta(p^0+|\vec{p}|)\big]
\frac{1}{e^{\beta |\vec{p}|}-1},
$$
no singularity could appear in integrand in Eq.(26) when
$|\vec{p}|\rightarrow 0$.  Hence we can conclude that
\begin{equation}
\Delta V_{2T}=0.
\end{equation}
In summary, from Eqs.(1),(11),(14),(16),(17),(18) (24) and (27) we
may write the final expression for the CJT effective potential for
quark propagator in the approximation $A(p^2)=1$ at finite
temperature and finite chemical potential in the real-time
formalism and in Landau gauge as follows.
\begin{equation}
V[G,G^*]=V_0+V_T
\end{equation}
\begin{eqnarray}
 V_0&=&V_{10}+V_{20}\nonumber \\
   &=&-2N_fN_c\Big\langle\ln\frac{\bar{p}^2+m^2(\bar{p}^2)}{\bar{p}^2}-
 \frac{2m^2(\bar{p}^2)}{\bar{p}^2+m^2(\bar{p}^2)}\Big\rangle_E \nonumber\\
  &&-6N_fN_cC_2(\underline{N}_c)\Big\langle\Big\langle
  g^2[\textrm{max}(\bar{p}^2,\bar{q}^2)]
  \frac{m(\bar{p}^2)}{\bar{p}^2+m^2(\bar{p}^2)}\cdot\frac{m(\bar{q}^2)}{\bar{q}^2+m^2(\bar{q}^2)}
  \cdot\frac{1}{(\bar{p}-\bar{q})^2}\Big\rangle\Big\rangle_E,
\end{eqnarray}
where $\langle\cdots \rangle_E$ and $\langle\langle\cdots
\rangle\rangle_E$ means the integrations of the Euclidean
momentums $\int d^4\bar{p}/(2\pi)^4$ and $\int
d^4\bar{p}d^4\bar{q}/(2\pi)^8$ respectively.
\begin{eqnarray}
  V_T &=& V_{1T}+V_{2T}\nonumber \\
   &=& -4\pi N_fN_c\Big\langle\sin^2\theta_p
  m^2(p^2)\delta\big[p^2-m^2(p^2)\big]\Big\rangle \nonumber\\
   && +i4\pi N_fN_cC_2(\underline{N}_c)g^2(0)\Big\langle\Big\langle\big[4m(p^2)m(q^2)-p^2-q^2\big]
 n(p^0-q^0)\delta\big[(p-q)^2\big] \nonumber \\
 && \left\{\frac{1}{p^2-m^2(p^2)+i\varepsilon}\cdot\frac{1}{q^2-m^2(q^2)+i\varepsilon}+
 i4\pi\sin^2\theta_p\delta\big[p^2-m^2(p^2)\big]\textrm{P}\frac{1}{q^2-m^2(q^2)}\right\}
 \Big\rangle\Big\rangle\nonumber \\
 &&-8\pi N_fN_cC_2(\underline{N}_c)\Big\langle\Big\langle\frac{g^2\big[(p-q)^2\big]}{(p-q)^2+i\varepsilon}
 \big[3m(p^2)m(q^2)+(p-q)^2 E(p,q)\big]\sin^2\theta_p\delta\big[p^2-m^2(p^2)\big]\nonumber \\
  &&\left\{\frac{i}{q^2-m^2(q^2)+i\varepsilon}-\pi
  \sin^2\theta_q\delta\big[q^2-m^2(q^2)\big]\right\}\Big\rangle\Big\rangle
\end{eqnarray}
In Eqs.(29) and (30), the running gauge coupling constant
$g^2\big[(p-q)^2\big]$ and the running quark mass function
$m(p^2)$ can be taken from the known results in zero temperature
QCD which are consistent with  RG analysis. After introducing the
infrared momentum cutoff $p_c^2$, we can write \cite{kn:11}
\begin{equation}
g^2\big[(p-q)^2\big]=2\pi^2A/\ln\frac{|(p-q)^2|+p_c^2}{\Lambda^2_{QCD}}
\end{equation}
and
\begin{equation}
m(p^2)=\hat{m}\left(\ln\frac{|p^2|+p_c^2}{\Lambda^2_{QCD}}\right)^{-A/2}+
\frac{2\pi^2A}{3}\frac{\phi}{(|p^2|+p_c^2)\varepsilon(p^2)}
\left(\ln\frac{|p^2|+p_c^2}{\Lambda^2_{QCD}}\right)^{A/2-1},\;\;
\varepsilon(p^2)\equiv\frac{p^2}{|p^2|},
\end{equation}
where $A=24/(33-2N_f)$, $\Lambda_{QCD}$ is the RG-invariant mass
scale parameter,
$\phi\equiv\langle0|\bar{\psi}\psi|0\rangle/(\ln\mu^2/\Lambda^2_{QCD})^{A/2}$
and $\hat{m}=m[\ln(\mu^2/\Lambda^2_{QCD})]^{A/2}$ are respectively
the RG-invariant quark-antiquark condensates and the current quark
mass, and $\mu$ is the renormalization scale parameter. Eqs.
(28)-(30) combined with Eqs. (31)-(32) give a calculable
expression for the thermal CJT effective potential for quark mass
function $m(p^2)$. Obviously, it is a function of temperature $T$,
quark chemical potential $\mu$, the quark-antiquark condensates
$\phi$ and, of course, the current quark mass $\hat{m}$ as well.
The result provides a useful means to research chiral symmetry
phase transition in QCD in the real-time formalism of thermal
field theory.

\end{document}